\documentclass[12pt]{iopart}
\usepackage{graphics}
\usepackage{epsfig}
\eqnobysec

\begin{document}

\title{Tensor polarization of deuterons passing through matter}

\author{V G Baryshevsky and A R Bartkevich}

\address{Research Institute for Nuclear Problems,
Belarusian State University, 11 Bobruiskaya Str., Minsk 220050,
Belarus} \ead{bar@inp.bsu.by; v\_baryshevsky@yahoo.com}

\begin{abstract}

It is shown that the magnitude of tensor polarization of the
deuteron beam, which arises owing to the spin dichroism effect,
depends appreciably on the angular width  of the detector that
registers the deuterons transmitted through the target. Even when
the angular width of the detector is much smaller than the mean
square angle of multiple Coulomb scattering, the beam's tensor
polarization depends noticeably on rescattering.  When the angular
width  of the detector is much larger than the mean square angle
of multiple Coulomb scattering (as well as than the characteristic
angle of elastic nuclear scattering), tensor polarization is
determined only by the total reaction cross sections for
deuteron-nucleus interaction, and elastic scattering processes
make no contribution to tensor polarization.

\end{abstract}

 \submitto{\JPG} \maketitle

\maketitle

\section{Introduction}
 The quasi-optical macroscopic
quantum phenomenon of birefringence, first described in
\cite{1,2}, combines the effects of particle spin rotation
(oscillation) and spin dichroism (i.e., the dependence of the
absorption coefficient on the particle spin state) and occurs when
high-energy particles with spin $S \ge 1$ traverse matter.  It is
analogous to the phenomenon known in optics as birefringence of
light in optically anisotropic media due to the dependence of the
refractive index  on the polarization state of light, i.e., on the
photon spin state. In contrast to light, whose wavelength largely
exceeds the interatomic distance in matter, the de Broglie
wavelength for a fast particle is much shorter than this distance.
According to \cite{1,2,3}, however, in this case one can also
introduce for particles a spin-dependent index of refraction.
Under such conditions, the birefringence effect for deuterons
takes place even in homogeneous, isotropic matter and results from
the intrinsic anisotropy of particles with spin $S \ge 1$.

An important implication of the spin dichroism effect is tensor
polarization arising in the initially unpolarized beam after it
has passed through an unpolarized target \cite{1,2}.

In 2005, the spin dichroism effect was first experimentally
observed for 5--20 MeV deuterons passing through a carbon target
\cite{4,5,6}. In 2007, tensor polarization arising in a nuclotron
extracted beam of initially unpolarized deuterons with a momentum
of 5\,GeV/c was measured \cite{7,8}.

The magnitudes of spin dichroism and tensor polarization of the
transmitted  deuterons are proportional to the difference between
the total interaction cross sections of the particle with the
nucleus for the states with magnetic quantum numbers $M=0$ and
$M=\pm1$ \cite{1,2,3}.

According to \cite{1,2,3,4,5,6,7,8},  this effect can be used not
only for studying the total spin--dependent interaction cross
section, but also for obtaining tensor-polarized deuteron beams.
Besides, this effect must be taken into account when interpreting
the results of planned experiments \cite{9} for measuring the
electric dipole moment of deuterons \cite{3,10}.

In theoretical description of this phenomenon, one should bear in
mind  that a deuteron colliding with a nucleus undergoes two
interactions: the Coulomb and nuclear ones.
The Coulomb interaction not only causes multiple Coulomb
scattering, but in some cases, it can also affect significantly
the magnitude of tensor polarization of the deuterons that have
passed through the target. It has been shown in \cite{11}, for
example, that the Coulomb--nuclear interference gives a clue to
the explanation of the experimentally observed fact that tensor
polarization reverses sign as the energy of  deuterons changes
from 5 to 20\,MeV.

Multiple scattering of deuterons in matter and the
scattering--angle dependence of the Coulomb--nuclear interference
account for the necessity to use the kinetic equation for the spin
density matrix in order to describe the spin--state dynamics of
the deuteron passing through matter \cite{12}.

 In the present paper, it is shown that tensor polarization of the
 deuterons escaping from the target in the direction of the in-coming primary beam
  depends on the collimator's angular width of
 the detector that registers the transmitted deuteron beam.
It is demonstrated that when  the angular width of the detector's
collimator is much larger than the characteristic angles of
deuteron scattering in the target, tensor polarization of
deuterons is determined only by the total cross sections of
inelastic processes in the interaction between the deuterons and
the nuclei.

This paper is arranged as follows. \Sref{s1} gives the kinetic
equation of the spin density matrix  and the spin--dependent
nuclear scattering amplitude. \Sref{s2.1} considers deuteron spin
dichroism in a thin target,where only single scattering is
essential.  \Sref{s2.2} discusses the effect that multiple
scattering makes on the magnitude of tensor polarization of the
deuteron beam moving in matter. A brief summary of the obtained
results is given in Conclusion.

\section{Kinetic equation for the spin density matrix}
\label{s1}

In order to describe the behavior of a beam of spin particles in
matter, let us introduce the spin density matrix $\hat{\rho}$ of
the system  "incident particle + target". The spin density matrix
$\hat {\rho }_{\rm{d}} $ of particles (deuterons, protons) moving
in matter is defined by the expression  $\hat {\rho }_\rmd =
\Tr_{\rm{T}} \hat {\rho }$, where $\Tr_{\rm{T}}$ means summation
of the  diagonal elements  of the matrix over all states of the
target. The equation for time evolution of the spin density matrix
$\hat {\rho }_{\rmd} $ has the form \cite{13}:

\[
\frac{\rmd\hat {\rho }_\rmd }{\rmd t} = - \frac{\,i}{\hbar }\left[
{\hat {H}_0,\hat {\rho }_\rmd } \right] + \mathop {\left(
{\frac{\partial \hat {\rho }_\rmd }{\partial t}}
\right)}\nolimits_{\rm{sct}},
\]
\noindent where the Hamiltonian $\hat {H}_0$ describes the
interaction between the particles and macroscopic external fields,
and the collision term $\mathop {\left( {\partial \hat {\rho
}_{\rm{d}} /
\partial t} \right)}\nolimits_{\rm{sct}} $ describes the  density matrix evolution
due to scattering by the target nuclei (atoms).

The explicit form of  $\mathop {\left( {\partial \hat {\rho
}_{\rm{d}} /
\partial t} \right)}\nolimits_{\rm{sct}} $ can be obtained when
considering scattering of the incident particle by the bound
target particle, provided that the following inequality holds:
$W_0 / \varepsilon _k \ll 1$. Here $W_0 $ is the binding energy of
the scatterer in the target, and $\varepsilon _k $ is the kinetic
energy of the incident particle. In this case, impulse
approximation applies to calculating the scattering process
\cite{14}. We shall further consider the case when the external
electric and magnetic fields are absent. This enables one to write
the below integro--differential representation of the kinetic
equation for the density matrix that describes the behavior of a
polarized particle beam as it passes through matter whose nuclei
(atoms) in the general case are polarized \cite{12}:
\begin{eqnarray}
 \fl\displaystyle
\frac{\rmd\hat{\rho}_{\rmd}(\vec{k},t)}{\rmd t}=\frac{2\pi
\rmi}{M_{\rm{r}}}N \Tr_{\rm{T}}\left(\hat{F}(\vec{k},\vec{k})
\hat{\rho}(\vec{k},t)-\hat{\rho}(\vec{k},t)\hat{F}^{\scriptscriptstyle+}(\vec{k},\vec{k})\right)\nonumber
\\
\displaystyle+\frac{N}{M_{\rm{r}}^{2}}\Tr_{\rm{T}}\int
\rmd^{3}\vec{k}'
\,\delta\left(\varepsilon_{k}-\varepsilon_{k'}-\frac{\vec{q}\,^{2}}{2M}\right)
\hat{F}(\vec{k},\vec{k}')\hat{\rho}(\vec{k}',t)
\hat{F}^{\scriptscriptstyle+}(\vec{k}',\vec{k}),\label{1.1}
\end{eqnarray}

\noindent where $\hat{F}(\vec{k},\vec{k}')$ is the amplitude of
particle scattering by the target nuclei, $N$ is the number of
target particles in  the unit volume, $M$ is the mass of the
scatterer, $M_{\rm{r}}$ is the reduced mass of the incident
particle and the nucleus, $\varepsilon_{k}$ is the kinetic energy
of the incident particle, $\vec {k}$ and $\vec {{k}'}$ are the
wave vectors of the beam particle before and after scattering,
respectively. The density matrix $\hat{\rho}(\vec{k})$ is the
direct product of the spin density matrix
$\hat{\rho}_{\rmd}(\vec{k})$ of the incident particle and the spin
density matrix $\hat{\rho}_{\rm{T}}(\vec{k})$ of the target
nuclei:
$\displaystyle\hat{\rho}(\vec{k})=\hat{\rho}_{\rmd}(\vec{k};\hat{\vec{S}}_{\rmd})\otimes\hat{\rho}_{\rm{T}}(\hat{\vec{S}}_{\rm{T}})$,
where $\hat{\vec{S}}_{\rmd}$ and $\hat{\vec{S}}_{\rm{T}}$ are the
spin operators of the beam particle and the scatterer,
respectively.

Kinetic equation  (\ref{1.1}) applies to such values of the
transmitted momentum $|\vec{q}|=|\vec {{k}'}-\vec {k}|$ that are
governed by the inequality  $q\gg r^{-1}$, where $r$ is the
correlation radius of the medium \cite{12}.

 \Eref{1.1} simplifies when a particle (proton, deuteron,
antiproton) passes through a target whose nuclei have a mass much
greater than the mass of the incoming particle. In this case, we
can neglect the effect of the incident particle's energy loss
through scattering. So we can neglect the recoil energy
$\displaystyle{\vec{q}\,^2}/{2M}$ in the $\delta$-function. As a
result, one obtains a simple kinetic equation describing the time
and spin evolution of the incident particle as it passes through
the target \cite{12,13}:
\begin{eqnarray} \fl\displaystyle
\frac{\rmd\hat{\rho}_{\rmd}(\vec{k},t)}{\rmd t}=\frac{2\pi \rmi
}{m}N \Tr_{\rm{T}}\left[\hat{F}(\vec{k},\vec{k})
\hat{\rho}(\vec{k},t)-\hat{\rho}(\vec{k},t)\hat{F}^{\scriptscriptstyle+}(\vec{k},\vec{k})\right]\nonumber
\\
\displaystyle+N\frac{k}{m}\Tr_{\rm{T}}\int \rmd\Omega_{\vec{k}'}
\hat{F}(\vec{k},\vec{k}')\hat{\rho}(\vec{k}',t)
\hat{F}^{\scriptscriptstyle+}(\vec{k}',\vec{k}),\label{1.2}
\end{eqnarray}

\noindent where $|\vec{k}|=|\vec{k}'|$ and $m$ is the mass of the
incident particle.

The first term on the right-hand side of (\ref{1.2}) can be
represented as follows:
\begin{eqnarray} \fl\displaystyle
\displaystyle\hat {F}(0)\hat {\rho }(\vec {k},t) - \hat
{\rho }(\vec {k},t)\hat {F}^{\scriptscriptstyle{+}}(0)\nonumber \\
=\left[ {\frac{1}{2}\left( {\hat {F}(0) + \hat
{F}^{{\scriptscriptstyle +}} (0)}\right),\hat {\rho }(\vec {k},t)}
\right] + \left\{ {\frac{1}{2}\left( {\hat {F}(0) - \hat
{F}^{\scriptscriptstyle +} (0)} \right),\hat {\rho }(\vec {k},t)}
\right\},\label{1.3}
\end{eqnarray}

\noindent where $\left[ , \right]$ is the commutator, $\left\{\, ,
\right\}$ is the anticommutator.

The part proportional to the commutator leads to rotation of the
polarization vector due to elastic coherent scattering \cite{13}
(as a result of the refraction effect \cite{1,2,3}); the
anticommutator describes the reduction in the intensity of the
transmitted beam.

The last term in (\ref{1.2}) determines the effect of incoherent
scattering on the change of $\hat{\rho}_d$ (in the general case,
the effect of single and multiple scattering).

For the sake of concreteness, let us consider the process of
deuteron passage through the target with spinless nuclei. In this
case, the density matrix $\hat {\rho}(\vec {k})$, as well as the
amplitude $\hat {F}(\vec {k},{\vec {k}}')$, contains only spin
variables of the scattered beam. For the amplitude $\hat {F}(\vec
{k},{\vec {k}}')$, we shall introduce the notation $\hat {f}(\vec
{k},{\vec {k}}')=\hat {F}(\vec{k},{\vec {k}}')$, where $\hat
{f}(\vec {k},{\vec {k}}') \equiv \hat {f}(\vec {k},{\vec
{k}}';\hat{\vec {S}}_{\rmd})$.

As a result, (ref{1.2}) can be written as follows:
\begin{eqnarray} \fl\displaystyle
\frac{\rmd\hat {\rho }_{\rmd}(\vec {k},z)}{\rmd z} = \frac{\pi
\rmi}{k}N\left[ {(\hat {f}(\vec {k},\vec {k}) + \hat
{f}^{\scriptscriptstyle +}(\vec {k},\vec {k})),\hat {\rho }_{\rmd
}(\vec {k},z)} \right] + \frac{\pi \rmi}{k}N\left\{ {(\hat
{f}(\vec {k},\vec {k}) - \hat {f}^{\scriptscriptstyle +}(\vec
{k},\vec {k})), \hat {\rho }_{\rmd} (\vec {k},z)}
\right\}\nonumber
\\
 + N\int {\rmd\Omega _{\vec {k}'} \hat {f}(\vec {k},{\vec {k}}') \hat
{\rho }_{\rmd} ({\vec {k}}',z)\hat {f}^{\scriptscriptstyle +}(\vec
{k}',{\vec {k}})}, \label{1.4}
 \end{eqnarray}
where $z=vt$ ($v$ is the particle velocity) is the distance
travelled by the incident particle in matter. Hereinafter, the
subscript $\rmd$ of the density matrix will be dropped.

In the case of deuterons (particles with spin 1), the polarization
state is characterized by the polarization vector $\vec {P}(\vec
{k}) =\Tr\hat {\rho }(\vec {k})\hat {\vec {S}}$ and the
quadrupolarization tensor (tensor of rank 2) ${\rm {\bf Q}}$,
whose components are defined as $Q_{ik} (\vec {k}) = \Tr\hat {\rho
}(\vec {k})\hat {Q}_{ik} $, where the operators  $\hat {Q}_{ik} $
can be represented in the form: $\hat {Q}_{ik} = \frac{3}{2}\left(
{\hat {S}_i \hat {S}_k + \hat {S}_k \hat {S}_i - \frac{4}{3}\delta
_{ik} {\hat{\mbox I}}} \right)$.

 The spin density matrix $\hat {\rho }(\vec {k})$ can be
written in the following general form:
\begin{eqnarray}
\hat {\rho }(\vec {k}) = \frac{1}{3}I(\vec {k}){\hat {\mbox I}} +
\frac{1}{2}\vec {P}(\vec {k})\hat {\vec {S}} + \frac{1}{9}Q_{ik}
(\vec {k})\hat {Q}_{ik},\label{1.5}
\end{eqnarray}

\noindent where $I(\vec {k}) = \Tr\hat {\rho }(\vec {k})$, ${\hat
{\mbox I}}$ is the $3\times 3$ identity matrix in the spin space.
Note that alongside with the quantities $\vec {P}$ and ${\rm {\bf
Q}}$, normalized spin characteristics of the beam are also used to
describe the beam polarization \cite{15, 16}.

For particles with spin 1 that are scattered by an unpolarized
nucleus, the amplitude $\hat {f}(\vec {k},{\vec {k}}')$  can be
expressed in terms of the deuteron spin operator $\hat {\vec
{S}}$, the quadrupolarization tensor $\hat{Q}_{ik} $, and the
combination of vectors $\vec {k}$ and ${\vec {k}}'$:
\begin{eqnarray} \displaystyle \hat {f}(\vec {k},{\vec {k}}') = A{\hat {I}} + B(\hat {\vec
{S}}\vec {\nu }) + C_1 \hat {Q}_{ik} \mu _i \mu _k + C_2 \hat
{Q}_{ik} \mu _{1i} \mu _{1k},\label{1.6}
\end{eqnarray}

\noindent where $A$, $B$, $C_1 $, and $C_2 $ are the parameters
depending on the scattering angle $\theta $; ${\vec {\nu } = [\vec
{k}\times {\vec {k}}'] / \vert [\vec {k}\times {\vec {k}}']\vert}
$;  $\vec {\mu } = (\vec {k} - {\vec {k}}') / \vert \vec {k} -
{\vec {k}}'\vert $, and $ \vec {\mu }_1 = (\vec {k} + {\vec {k}}')
/ \vert \vec {k} + {\vec {k}}'\vert $.

In view of (\ref{1.6}),  for the  zero-angle scattering amplitude
$\hat {f}(\vec {k},\vec {k})$, we have
\begin{eqnarray} \hat {f}(\vec {k},\vec {k}) = f_0 (0) + f_1 (0)(\vec {S}\vec
{n})^2,\label{1.7}
\end{eqnarray}
\noindent where $\vec {n} = \vec {k} / k$ is the unit vector in
the direction  of $\vec {k}$, and the following notation is
introduced: $f_0 = A - 2C_1 - 2C_2 $, $f_1 = 3C_2 $.

In the general case, $f_0 $ and $f_1 $ are complex functions, and
according to the optical theorem, the imaginary parts  of $f_0 $
and $f_1 $ can be expressed in terms of the corresponding total
cross sections, namely,
\begin{eqnarray}
\displaystyle \mathrm{Im} f_0 (0) = \frac{k}{4\pi }\,\sigma
_{\rm{tot}}^0, \qquad\mathrm{Im}f_1 (0) = \frac{k}{4\pi }\left[
{\sigma _{\rm{tot}}^{\pm 1} - \sigma _{\rm{tot}}^0 }
\right],\label{1.8}
\end{eqnarray}
\noindent where $\sigma _{\rm{tot}}^0 $, $\sigma _{\rm{tot}}^{\pm
1} $ are the total scattering cross sections for the initial spin
state of the deuteron with magnetic quantum numbers $M = 0$ and $M
= \pm 1$, respectively (the quantization axis $z$ is directed
along $\vec {n}$).

Deuterons interact with the target nuclei via the Coulomb and
nuclear  interactions. The  amplitude $\hat {f}(\vec {k},{\vec
{k}}')$ of deuteron scattering by the target nuclei in this case
can be represented in the form \cite{12,17}:
\begin{eqnarray} \hat {f}(\vec {k},{\vec {k}}') = \hat {f}_{\rm{coul}} (\vec {k},{\vec
{k}}') + \hat {f}_{\rm{nucl,coul}} (\vec {k},{\vec
{k}}'),\label{1.9}
\end{eqnarray}
\noindent where $\hat {f}_{\rm{coul}}$ is the amplitude of
Coulomb scattering of the deuteron by the
 nucleus in the absence of nuclear interaction, $\hat {f}_{\rm{nucl,coul}} $ is the amplitude of scattering of
 Coulomb--distorted waves by the nuclear potential.
\section{The phenomenon of spin dichroism}
\label{s2}

The phenomenon of deuteron spin dichroism (i.e., the dependence of
the absorption coefficient on the spin state of the particle
moving in matter) leads to the appearance of tensor polarization
in the initially unpolarized beam transmitted through the
unpolarized target.

The unpolarized deuteron beam can be considered as an incoherent
mixture of three polarized beams having the same intensity and
spin states with magnetic quantum numbers $M=-1,\,0$, and $+1$,
respectively (the quantization axis is directed parallel to the
deuteron momentum).  The total  scattering cross section
$\sigma_{\rm{tot}}$ of deuterons by a nucleus depends on the
polarization state of the incident particle; namely, in the
interaction with an unpolarized nucleus,
$\sigma_{\rm{tot}}^{+1}=\sigma_{\rm{tot}}^{-1}\neq\sigma_{\rm{tot}}^{0}$.
As a result, the partial intensities
$I^{\scriptscriptstyle(\pm1)}$ and $I^{\scriptscriptstyle(0)}$ of
the beam transmitted through the target appear to be unequal, and
so the spin dichroism phenomenon can be characterized by the
parameter
\begin{eqnarray}
D=\frac{I^{\scriptscriptstyle(\pm1)}-I^{\scriptscriptstyle(0)}}{I^{\scriptscriptstyle(\pm1)}+I^{\scriptscriptstyle(0)}},\label{2.1}
\end{eqnarray}

\noindent where $I^{\scriptscriptstyle(0)}$,
$I^{\scriptscriptstyle(\pm1)}$ are the intensities of the deuteron
beam transmitted through the unpolarized target for the case when
the initial state of the deuteron is defined by quantum numbers
$M=0$ and  $M=\pm1$, respectively.

Spin dichroism leads to the appearance of tensor polarization in
the transmitted beam. According to the definition (see e.g.,
\cite{15,16}), the diagonal components of the normalized tensor
polarization can be written in the form:
\begin{eqnarray}
p_{xx}=p_{yy}=-\frac{1}{2}\frac{I^{\scriptscriptstyle(+1)}+I^{\scriptscriptstyle(-1)}-2I^{\scriptscriptstyle(0)}}
{I^{\scriptscriptstyle(+1)}+I^{\scriptscriptstyle(-1)}+I^{\scriptscriptstyle(0)}},\qquad
p_{zz}=\frac{I^{\scriptscriptstyle(+1)}+I^{\scriptscriptstyle(-1)}-2I^{\scriptscriptstyle(0)}}
{I^{\scriptscriptstyle(+1)}+I^{\scriptscriptstyle(-1)}+I^{\scriptscriptstyle(0)}},\label{2.2}
\end{eqnarray}

\noindent where $p_{xx}=Q_{xx}/I$, $p_{yy}=Q_{yy}/I$, and
$p_{zz}=Q_{zz}/I$.

In the case when the beam passes through the unpolarized target
that is considered here, the intensities
$I^{\scriptscriptstyle(+1)}=I^{\scriptscriptstyle(-1)}$. As a
result, we have the following expression for  $p_{\mathrm{zz}}$:
\begin{eqnarray}
p_{zz}=2\frac{I^{\scriptscriptstyle(\pm1)}-I^{\scriptscriptstyle(0)}}{I},\label{2.3}
\end{eqnarray}
where $I$ is the intensity of the initially unpolarized beam,
$I=I^{\scriptscriptstyle(+1)}+I^{\scriptscriptstyle(-1)}+I^{\scriptscriptstyle(0)}
=2I^{\scriptscriptstyle(\pm1)}+I^{\scriptscriptstyle(0)}$.

\subsection{Tensor polarization of deuterons passing
through a thin target} \label{s2.1}

Let us consider particle transmission through the target with
thickness $z$, which is smaller than the deuteron's free path in
matter, i.e., $z < 1 / N\sigma $, $\sigma $ is the total
scattering cross section of the deuteron by the nucleus. In the
first--order perturbation theory, the solution of kinetic equation
(\ref{1.4}) can be represented in the form:
\begin{eqnarray}\fl
\displaystyle\hat {\rho }(\vec {k},z) = \hat {\rho }(\vec {k},0) +
\frac{\pi \rmi}{k}N\left[ {(\hat {f}(\vec {k},\vec {k}) + \hat
{f}^{\scriptscriptstyle+} (\vec {k},\vec {k})),\hat {\rho }(\vec
{k},0)} \right]z\nonumber
 \\
\displaystyle+\frac{\pi \rmi}{k}N\left\{ {(\hat {f}(\vec {k},\vec
{k}) - \hat {f}^{\scriptscriptstyle+}(\vec {k},\vec {k})),\hat
{\rho }(\vec {k},0)} \right\}z
 + N\hat {f}(\vec {k},\vec {k}_0 )\hat {\rho }(0)\hat {f}^{\scriptscriptstyle+}(\vec {k}_0
,\vec {k})z,\label{2.4}
\end{eqnarray}

\noindent where $\hat {\rho }(\vec {k},0)$ is the density matrix
of the beam at entering the target, i.e., when $z = 0$. It
describes the momentum distribution of the incoming particles with
respect to the direction  $\vec {k}_0$.

In deriving (\ref{2.4}), it was also assumed that the beam's
initial angular distribution is much smaller than the
characteristic angular width of the differential scattering cross
section. In this case, the amplitude $\hat {f}(\vec {k},{\vec
{k}}')$ in the term $\displaystyle\int {\rmd\Omega _{\vec {k}'}
\hat {f}(\vec {k},{\vec {k}}')\hat {\rho }({\vec {k}}',0)\hat
{f}^{\scriptscriptstyle+}({\vec {k}}',\vec {k})} $ can be removed
from the integral sign  at point ${\vec {k}}'=\vec {k}_0$. As a
result, we have $\displaystyle\int {\rmd\Omega _{\vec {k}'} \hat
{f}(\vec {k},{\vec {k}}')\hat {\rho }({\vec {k}}',0)\hat
{f}^{\scriptscriptstyle+}({\vec {k}}',\vec {k})} \simeq \hat
{f}(\vec {k},\vec {k}_0 )\hat {\rho }(0)\hat
{f}^{\scriptscriptstyle+}(\vec {k}_0 ,\vec {k})$, where
$\displaystyle\hat {\rho }(0) = \int{\rmd\Omega _{\vec {k}'}\hat
{\rho }({\vec {k}}',0)} $ is the spin part of the beam's density
matrix $\hat {\rho }(\vec {k},0)$. The term $N\displaystyle\int
{\rmd\Omega _{\vec {k}'}} \hat {f}(\vec {k},{\vec {k}}')\hat {\rho
}({\vec {k}}',0)\hat {f}^{\scriptscriptstyle+}({\vec {k}}',\vec
{k})$ or $N\hat {f}(\vec {k},\vec {k}_0 )\hat {\rho }(0)\hat
{f}^{\scriptscriptstyle+}(\vec {k}_0 ,\vec {k})$ describes the
contribution to the evolution of the density matrix due to
 single scattering of particles in the direction of vector $\vec {k}$.

\begin{figure}[!h]
\centering
\includegraphics[scale=0.7]{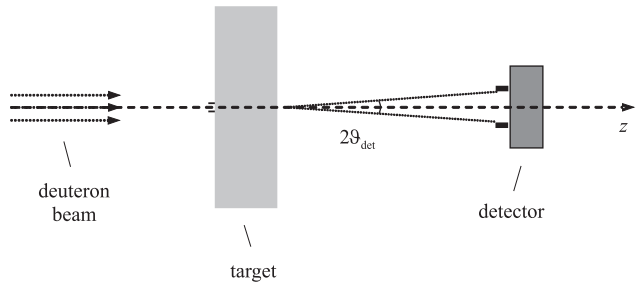}
\caption{Scheme of detecting the transmitted beam.}\label{f1}
\end{figure}

Using the solution  (\ref{2.4}) and the explicit form of the spin
structure of the amplitude $\hat {f}(\vec {k},{\vec {k}}')$ [see
(\ref{1.6})], one can find the dependence of the beam's intensity
and polarization characteristics on the direction of particle
scattering and on the distance $z$ passed by the deuteron in
matter. In a real experiment, the collimator of the detector has a
finite angular width, and so  the scattered particles are
registered within a certain range of finite momenta. For this
reason, the characteristics of the beam transmitted through the
target should be studied in the range of solid angles $\Delta
\Omega $ with respect to the initial direction of
 beam propagation (see \fref{f1}).  If the collimator has axial
symmetry, then $\Delta \Omega $ is defined by the angular width
$2\vartheta _{\rm{det}} $ of the detector's collimator (for
further calculations, it is also assumed that $\vartheta
_{\rm{det}} $ is much larger than the initial angular distribution
of the beam).

We shall further consider the case when the unpolarized beam is
incident onto the target.

 The number of particles $\mathcal{I}^{\scriptscriptstyle(0)}$, $\mathcal{I}^{\scriptscriptstyle(\pm1)}$
 with quantum numbers $M=0$ and $M=\pm1$, which are
registered by the detector with angular width $\Delta \Omega $ is
determined by the diagonal matrix elements $\displaystyle\langle
\Psi_{\scriptscriptstyle M=0}|\int_{\Delta \Omega}\rmd\Omega
_{\vec {k}}\hat{\rho}(\vec {k},z)|\Psi_{\scriptscriptstyle
M=0}\rangle$ and $\displaystyle\langle \Psi_{\scriptscriptstyle
M=\pm1}|\int_{\Delta \Omega}\rmd\Omega _{\vec {k}}\hat{\rho}(\vec
{k},z)|\Psi_{\scriptscriptstyle M=\pm1}\rangle$ of the spin
density matrix (the quantization axis is chosen along the
direction $\vec{n}=\vec{k}/k$):
\begin{eqnarray}
\displaystyle\mathcal{I}^{\scriptscriptstyle(0)}(z)&=&I_{0}^{\scriptscriptstyle(0)}\left[1-\displaystyle
N\sigma^{\scriptscriptstyle0}_{\rm{tot}}z+
N\int_{\scriptscriptstyle\Delta\Omega}\rmd\Omega\,\langle
\Psi_{\scriptscriptstyle
M=0}|\left(\frac{\rmd\hat{\sigma}}{\rmd\Omega}\right)_{\rm{sc}}|\Psi_{\scriptscriptstyle
M=0}\rangle z\right],\nonumber\\
\displaystyle\mathcal{I}^{\scriptscriptstyle(\pm1)}(z)&=&I_{0}^{\scriptscriptstyle(\pm1)}\left[1-\displaystyle
N\sigma^{\scriptscriptstyle\pm1}_{\rm{tot}}z+
N\int_{\scriptscriptstyle\Delta\Omega}\rmd\Omega\,\langle
\Psi_{\scriptscriptstyle
M=\pm1}|\left(\frac{\rmd\hat{\sigma}}{\rmd\Omega}\right)_{\rm{sc}}|\Psi_{\scriptscriptstyle
M=\pm1}\rangle\, z\right],\label{2.5}
\end{eqnarray}

\noindent where $\displaystyle{\cal I} \equiv \int_
{\scriptscriptstyle\Delta \Omega}\rmd\Omega I(\vec{k},z)$;
$I_{0}^{\scriptscriptstyle(0)}$ and
$I_{0}^{\scriptscriptstyle(\pm1)}$ are the partial values of the
flow of the deuteron beam at ${z=0}$ for the spin states
$|\Psi_{\scriptscriptstyle M=0}\rangle$ and
$|\Psi_{\scriptscriptstyle M=\pm1}\rangle$, respectively. For the
case of unpolarized beams, considered below,
$I_{0}^{\scriptscriptstyle(0)}=\frac{1}{3}I_{0}$  and
$I_{0}^{\scriptscriptstyle(\pm1)}=\frac{1}{3}I_{0}$, where $I_{0}$
is the deuteron flow at the target entrance; $\displaystyle I_{0}
= \int {\rmd\Omega I(\vec {k},0)} \,$. The differential scattering
cross section of the deuteron is
$\left(\rmd\hat{\sigma}/\rmd\Omega\right)_{\rm{sc}}$.

To find tensor polarization of the registered beam, let us make
use of  (\ref{2.3}) and (\ref{2.5}). As a result, one can obtain
the following expression for $p_{zz}$:
\begin{eqnarray}\fl
p_{zz}\simeq
-\frac{2}{3}N\left(\sigma^{\scriptscriptstyle\pm1}_{\rm{tot}}-\sigma^{\scriptscriptstyle0}_{\rm{tot}}\right)z\nonumber
\\
+\frac{2}{3}N\, \int_{\scriptscriptstyle\Delta\Omega}\rmd\Omega
\,\left\{ \langle \Psi_{\scriptscriptstyle
M=\pm1}|\left(\frac{\rmd\hat{\sigma}}{\rmd\Omega}\right)_{\rm{sc}}|\Psi_{\scriptscriptstyle
M=\pm1}\rangle\right.\nonumber
\\
 -\left. \langle \Psi_{\scriptscriptstyle
M=0}|\left(\frac{\rmd\hat{\sigma}}{\rmd\Omega}\right)_{\rm{sc}}|\Psi_{\scriptscriptstyle
M=0}\rangle\,\right\} z, \label{2.6}
\end{eqnarray}
that is,
\begin{eqnarray}\fl
p_{zz}\simeq
-\frac{2}{3}N\left(\sigma^{\scriptscriptstyle\pm1}_{\rm{r}}-\sigma^{\scriptscriptstyle0}_{\rm{r}}\right)z\nonumber
\\
-\frac{2}{3}N\,\left[(\sigma^{\scriptscriptstyle\pm1}_{\rm{el}}-\sigma^{\scriptscriptstyle0}_{\rm{el}})-
\int_{\scriptscriptstyle\Delta\Omega}\rmd\Omega\, \left\{ \langle
\Psi_{\scriptscriptstyle
M=\pm1}|\left(\frac{\rmd\hat{\sigma}}{\rmd\Omega}\right)_{\rm{sc}}|\Psi_{\scriptscriptstyle
M=\pm1}\rangle\right.\right. \nonumber\\- \left.\left. \langle
\Psi_{\scriptscriptstyle
M=0}|\left(\frac{\rmd\hat{\sigma}}{\rmd\Omega}\right)_{\rm{sc}}|\Psi_{\scriptscriptstyle
M=0}\rangle\,\right\}\right]\, z, \label{2.7}
\end{eqnarray}
where  $\sigma_{\rm{r}}$ is the inelastic part of the total cross
section, which includes all possible nuclear reactions,
$\sigma_{\rm{el}}$ is the  total cross section of elastic
scattering.

 As
is seen, taking account of the scattered particles leads to the
presence in the tensor polarization of the term depending on the
angular width of the detector's collimator. Particularly, if
$\vartheta _{\rm{det}}\rightarrow 0$, then
$\displaystyle\int_{\scriptscriptstyle\Delta\Omega}\rmd\Omega
\left(\frac{\rmd\hat{\sigma}}{\rmd\Omega}\right)_{\rm{sc}}\rightarrow
0$. As a consequence [see (\ref{2.6})], tensor polarization
depends on the difference between the total cross section for the
deuterons  in the initial spin state  with $M=0$ and that with
$M=\pm 1$:
\begin{eqnarray}
p_{zz}(\vartheta _{\rm{det}}\rightarrow 0)\simeq
-\frac{2}{3}N\left(\sigma^{\scriptscriptstyle\pm1}_{\rm{tot}}-\sigma^{\scriptscriptstyle0}_{\rm{tot}}\right)z.
\label{2.8}
\end{eqnarray}

If $\vartheta _{\rm{det}}\rightarrow\pi$, then all the scattered
particles are registered. In this case, the second term is
proportional to the difference between the total cross sections of
elastic scattering:
\begin{eqnarray}
\sigma^{\scriptscriptstyle\pm1}_{\rm{el}}&=&\int\rmd\Omega \langle
\Psi_{\scriptscriptstyle
M=\pm1}|\left(\frac{\rmd\hat{\sigma}}{\rmd\Omega}\right)_{\rm{sc}}|\Psi_{\scriptscriptstyle
M=\pm1}\rangle,\nonumber\\
\sigma^{\scriptscriptstyle 0}_{\rm{el}}&=&\int\rmd\Omega \langle
\Psi_{\scriptscriptstyle
M=0}|\left(\frac{\rmd\hat{\sigma}}{\rmd\Omega}\right)_{\rm{sc}}|\Psi_{\scriptscriptstyle
M=0}\rangle.\label{2.9}
\end{eqnarray}

As a result, according to (\ref{2.7}), tensor polarization
$p_{zz}$ is described by the following expression:
\begin{eqnarray}
p_{zz}(\vartheta _{\mathrm{det}}=\pi)\simeq
-\frac{2}{3}N\left(\sigma^{\scriptscriptstyle\pm1}_{\rm{r}}-\sigma^{\scriptscriptstyle0}_{\rm{r}}\right)z.
\label{2.10}
\end{eqnarray}

The differential  cross section of elastic scattering for fast
particles scattered by the nucleus  achieves the largest values in
the range of angles $\theta\leq 1/kR$ ($R$ is the radius of action
of forces) and decreases rapidly  with growing  $\theta$
($\theta\gg 1/kR$).  For this reason,  $p_{zz}$ achieves the value
(\ref{2.10}) already when $\vartheta_{\rm{det}}\gg 1/kR$.

Thus, in the range of large values of $\vartheta_{\rm{det}}$,
tensor polarization only depends on the difference between the
inelastic scattering cross sections for deuteron states with $M=0$
and $M=\pm 1$. It is quite understandable because in this case,
there is no elastic-scattering-related losses in the flow of
particles registered by the detector: the particles that have
passed through the target without being scattered and those that
have been elastically scattered equally  get into the detector.

\subsection{Influence of multiple Coulomb scattering on tensor
polarization of deuterons passing through matter}\label{s2.2}

The solution (\ref{2.4}) applies to such target thicknesses $z$,
for which multiple scattering in matter can be neglected, i.e.,
for $z \le 1 / N\sigma _{\rm{coul}} $.

To solve (\ref{1.4})  for the case $z>1/N\sigma_{\rm{coul}}$, let
us take into account that  in  the considered energy range,
$\sigma_{\rm{nucl}}\ll \sigma_{\rm{coul}}$. As a result, for the
zeroth approximation we use the solution of kinetic equation
(\ref{1.4}), where the collision term is only determined by the
the Coulomb interaction between the incident particle and the
nuclei of matter. The contribution coming to the evolution of the
spin density matrix from nuclear scattering and the
Coulomb-nuclear interference can be considered as a correction,
which is true  for the cases when $z\leq 1 / N\sigma_{\rm{nucl}}
$.

 \subsubsection{Solution of the Kinetic Equation.}

Using the optical theorem (\ref{1.8}),  equation (\ref{1.4}) can
be written as follows:
\begin{eqnarray}\fl
\frac{\rmd\hat {\rho }(\vec {k},z)}{\rmd z} = - N\sigma
_{\rm{tot}}^{\scriptscriptstyle0} \hat {\rho }(\vec {k},z) -
\frac{N}{2}\left( {\sigma _{\rm{tot}}^{\scriptscriptstyle\pm 1} -
\sigma _{\rm{tot}}^{\scriptscriptstyle0} } \right)\left\{
{(\vec{S}\vec{n})^{2},\hat {\rho }(\vec {k},z)} \right\}\nonumber
\\
+\frac{2\pi \rmi}{k}N \mathrm{Re} f_1 (0)\left[
{(\vec{S}\vec{n})^{2},\hat {\rho }(\vec {k},z)} \right]+ N\int
{\rmd\Omega _{\vec {k}'} \hat {f}(\vec {k},{\vec {k}}')\hat {\rho
}({\vec {k}}',z)\hat {f}^{\scriptscriptstyle +}(\vec {k}',{\vec
{k}})}\label{2.11}
 \end{eqnarray}

\noindent where $\vec{n}$ is the unit vector in the direction of
the momentum $\vec{k}$.

It follows from the form of the amplitude (\ref{1.9}) that the
elastic part of the total cross section
 $\sigma _{\rm{el}}^{\scriptscriptstyle
0,\pm 1} $  can be presented as a sum of the terms describing the
Coulomb scattering cross section, the nuclear cross section, and
the contributions to the cross section coming from the
interference  between the Coulomb and nuclear interactions:
$\sigma _{\rm{el}}^{\scriptscriptstyle 0,\pm 1} = \sigma
_{\rm{coul}}^{\scriptscriptstyle 0,\pm 1} + \sigma
_{\rm{nucl,coul}}^{\scriptscriptstyle 0,\pm 1} + \sigma
_{\rm{nucl}}^{\scriptscriptstyle 0,\pm 1}$. Let us introduce the
following notations: $\sigma _{\rm{NC}}^{\scriptscriptstyle 0,\pm
1} = \sigma _{\rm{nucl,coul}}^{\scriptscriptstyle 0,\pm 1} +
\sigma _{\rm{nucl}}^{\scriptscriptstyle 0,\pm
1}+\sigma_{\rm{r}}^{\scriptscriptstyle 0,\pm 1}$. Thus, the cross
section  $\sigma _{\rm{tot}}^{\scriptscriptstyle 0,\pm 1} $ can be
presented in the form:
\begin{eqnarray}
\sigma _{\rm{tot}}^{\scriptscriptstyle 0,\pm 1} = \sigma
_{\rm{coul}}^{\scriptscriptstyle 0,\pm 1} + \sigma
_{\rm{NC}}^{\scriptscriptstyle 0,\pm 1}.\label{2.12}
 \end{eqnarray}

In view of (\ref{2.12}) and the representation (\ref{1.9})  for
the scattering amplitude, kinetic equation (\ref{2.11}) can be
written in the form
\begin{eqnarray}\fl
\frac{\rmd\hat {\rho }(\vec {k},z)}{\rmd z} = - N\sigma
_{\rm{coul}}^{\scriptscriptstyle 0} \hat {\rho }(\vec {k},z) -
\frac{N}{2}\left( {\sigma _{\rm{coul}}^{\scriptscriptstyle \pm 1}
- \sigma _{\rm{coul}}^{\scriptscriptstyle 0} } \right)\left\{
{(\vec{S}\vec{n})^{2},\hat {\rho }(\vec {k},z)} \right\} \nonumber
\\
+\frac{2\pi \rmi}{k}N \mathrm{Re} f_1 ^{\rm{coul}}(0)\left[
{(\vec{S}\vec{n})^{2},\hat {\rho }(\vec {k},z)} \right] \nonumber
\\+ N\int
{\rmd\Omega _{\vec {k}'} \hat {f}_{\rm{coul}} (\vec {k},{\vec
{k}}')\hat {\rho }({\vec {k}}',z)\hat {f}^{\scriptscriptstyle
+}_{\rm{coul}} ({\vec {k}}',\vec {k})}\nonumber
\\
- N\sigma _{\rm{NC}}^{\scriptscriptstyle 0} \hat {\rho }(\vec
{k},z) - \frac{N}{2}\left( {\sigma
_{\rm{NC}}^{\scriptscriptstyle\pm 1} - \sigma
_{\rm{NC}}^{\scriptscriptstyle0} } \right)\left\{
{(\vec{S}\vec{n})^{2},\hat {\rho }(\vec {k},z)} \right\}\nonumber
\\ +
\frac{2\pi \rmi}{k}N \mathrm{Re} f_1 ^{\rm{nucl}}(0)\left[
{(\vec{S}\vec{n})^{2},\hat {\rho }(\vec {k},z)} \right]\nonumber
\\
+ N\int \rmd\Omega_{\vec {k}'} \left(\hat {f}_{\rm{coul}} (\vec
{k},{\vec {k}}')\hat {\rho }({\vec {k}}',z)\hat
{f}^{\scriptscriptstyle +}_{\rm{nucl,coul}} ({\vec {k}}',\vec
{k})\right.\nonumber
\\
+\left.\hat {f}_{\rm{nucl,coul}} (\vec {k},{\vec {k}}')\hat {\rho
}({\vec {k}}',z)\hat {f}^{\scriptscriptstyle +}_{\rm{coul}} ({\vec
{k}}',\vec {k}) \right)\nonumber
\\
 \displaystyle
 + N\int {\rmd\Omega _{\vec {k}'} \hat {f}_{\rm{nucl,coul}} (\vec {k},{\vec
{k}}')\hat {\rho }({\vec {k}}',z)\hat {f}^{\scriptscriptstyle
+}_{\rm{nucl,coul}} ({\vec {k}}',\vec {k}).}\label{2.13}
 \end{eqnarray}

As stated above, the solution of (\ref{2.13}) can be presented as
follows:
\begin{eqnarray}\hat {\rho }(\vec {k},z) = \hat {\rho }^{(0)}(\vec {k},z) + \hat
{\rho }^{(1)}(\vec {k},z) + ...,\label{2.14}
\end{eqnarray}

\noindent where $\hat {\rho }^{(0)}(\vec {k},z)$ is the zeroth
approximation, which is the solution of kinetic equation
(\ref{2.13}) in the case when only the Coulomb interaction between
the deuteron beam and the nuclei of matter is taken into account;
$\hat {\rho }^{(1)}(\vec {k},z)$ is the first-order perturbation
theory correction, which includes the contribution of nuclear
scattering to the evolution of the beam's polarization
characteristics.

The equation for  $\hat {\rho }^{(0)}(\vec {k},z)$ has the form:
\begin{eqnarray}\fl
\frac{\rmd\hat {\rho }^{(0)}(\vec {k},z)}{\rmd z} = - N\sigma
_{\rm{coul}}^{\scriptscriptstyle 0} \hat {\rho }^{(0)}(\vec {k},z)
- \frac{N}{2}\left( {\sigma _{\rm{coul}}^{\scriptscriptstyle\pm 1}
- \sigma _{\rm{coul}}^{\scriptscriptstyle 0} } \right)\left\{
{(\vec{S}\vec{n})^{2},\hat {\rho }^{(0)}(\vec {k},z)}
\right\}\nonumber
\\
+\frac{2\pi \rmi}{k}N \mathrm{Re} f_1 ^{\rm{coul}}(0)\left[
{(\vec{S}\vec{n})^{2},\hat {\rho }^{(0)}(\vec {k},z)}
\right]\nonumber
\\
+ N\int {\rmd\Omega _{\vec {k}'} \hat {f}_{\rm{coul}} (\vec
{k},{\vec {k}}')\hat {\rho }^{(0)}({\vec {k}}',z)\hat
{f}^{\scriptscriptstyle +}_{\rm{coul}} (\vec {k},{\vec {k}}')} ,
\label{2.15}
 \end{eqnarray}

\noindent where the spin structure of $\hat {f}_{\rm{coul}} (\vec
{k},{\vec {k}}')$ in the general case has the form  (\ref{1.6}).

In the small-angle approximation, the terms in (\ref{1.6}) that
are proportional to $B$, $C_1 $, and   $C_2 $ in the Coulomb
amplitude, for ${\vec {k}}'\neq \vec {k}$ lead  to depolarization
of the registered beam. The magnitude of this depolarization is
determined by the quantity $b_{\rm{g}}^2 \overline {\theta ^2} z$
($\displaystyle b_{\rm{g}} = \frac{g - 2}{2}\frac{\gamma ^2 -
1}{\gamma } + \frac{\gamma - 1}{\gamma }$, $g$ is the gyromagnetic
ratio, and $\displaystyle \overline {\theta ^2}$ is the mean
square angle of Coulomb scattering per unit length) \cite{3,18}.
For the considered path lengths, which are of the order of the
nuclear collision length or smaller,  the degree of depolarization
for deuterons having energies in the range of hundreds of
megaelectron-volts and moving in carbon is $\sim 10^{-3}$. As is
seen, deuteron depolarization is insignificant and will further be
neglected. As a result, in the case of Coulomb interaction, it is
sufficient to consider the spin-independent part of scattering
amplitude. Then (\ref{2.15}) reads:
\begin{eqnarray}
\frac{\rmd\hat {\rho }^{(0)}(\vec {k},z)}{\rmd z} = - N\sigma
_{\rm{coul}}^{\scriptscriptstyle0} \hat {\rho }^{(0)}(\vec {k},z)
+ N\int {\rmd\Omega _{\vec {k}'} \left| {a(\vec {k},{\vec {k}}')}
\right|^2\hat {\rho }^{(0)}({\vec {k}}',z)},\label{2.16}
\end{eqnarray}

\noindent where $a(\vec {k},{\vec {k}}')$ denotes the spinless
part of the amplitude  $\hat {f}_{\rm{coul}} (\vec {k},{\vec
{k}}')$.

Expression (\ref{2.16}) is an integro-differential equation. In
the limit of small scattering angles, it can be solved by
expanding the function $\hat {\rho }^{(0)}({\vec {k}}')$ in terms
of a small parameter $\vec {q}={\vec{k}}'-\vec{k} \ll \vec{k}$
(transmitted momentum) \cite{19}:
\begin{eqnarray}\fl
\displaystyle\hat {\rho }^{(0)}({\vec {k}}',z) \approx \hat {\rho
}^{(0)}(\vec {k},z) + \frac{\partial \hat {\rho }^{(0)}}{\partial
k_x }q_x + \frac{\partial \hat {\rho }^{(0)}}{\partial k_y }q_y
\nonumber
\\
+\frac{1}{2}\frac{\partial ^2\hat {\rho }^{(0)}}{\partial k_x
^2}q_x ^2 + \frac{\partial ^2\hat {\rho }^{(0)}}{\partial k_x
\partial k_y }q_x q_y + \frac{1}{2}\frac{\partial ^2\hat {\rho
}^{(0)}}{\partial k_y ^2}q_y ^2 + ...,\label{2.17}
\end{eqnarray}

\noindent where $q_x = q\cos \varphi $, $q_y = q\sin \varphi $,
and $q \simeq k\theta $ with $\theta $ being the scattering angle
(the angle between vectors  $\vec {k}$ and $\vec {k}'$).

Substitution of the expansion (\ref{2.17}) into (\ref{2.16}) and
integration over the azimuthal angle  $\varphi $ gives
\begin{eqnarray}
\frac{\rmd\hat {\rho }^{(0)}(\vec {k},z)}{\rmd z} =
\frac{\overline {\,\theta ^2} }{4}\Delta \hat {\rho }^{(0)}(\vec
{k},z) + \frac{\overline {\,\theta ^4} }{64}\Delta \Delta \hat
{\rho }^{(0)}(\vec {k},z) + ...,\label{2.18}
\end{eqnarray}

\noindent the operator $\Delta $ affects the transversal
components of vector $\vec {k}$: $\displaystyle\Delta =
\frac{\partial ^2}{\partial n_x ^2} + \frac{\partial ^2}{\partial
n_y ^2}$, where $\vec {n} = \vec {k} / k$, $\displaystyle\overline
{\,\theta ^2} = N\int {\theta ^2} \frac{\rmd\sigma _{\rm{coul}}
}{\rmd\Omega }\rmd\Omega $, and $\displaystyle\overline {\,\theta
^4} = N\int {\theta ^4} \frac{\rmd\sigma _{\rm{coul}} }{\rmd\Omega
}\rmd\Omega $, etc.

If we confine ourselves to the first term  on the right-hand side
of (\ref{2.18}), then integro-differential equation (\ref{2.17})
can be reduced to  a  differential equation. The limiting angle in
such approximation is obtained from the condition
\begin{eqnarray}
\theta _{\max }^2 < \frac{16\Delta {\rho }^{(0)}(\vec
{k},z)}{\Delta \Delta  {\rho }^{(0)}(\vec {k},z)}.\label{2.19}
\end{eqnarray}

\noindent As a result, we have the equation for the spin density
matrix $\hat{\rho}^{(0)}(\vec{k}, z)$, which coincides in form
with the equation for the spin density matrix describing multiple
scattering of spinless particles:
\begin{eqnarray}
\frac{\rmd\hat {\rho }^{(0)}(\vec {k},z)}{\rmd z} =
\frac{\overline {\,\theta ^2} }{4}\Delta \hat {\rho }^{(0)}(\vec
{k},z).\label{2.20}
\end{eqnarray}

\noindent For the initial condition $\hat {\rho }^{(0)}(\vec {k},z
= 0) = \hat {\rho }_0 \delta (n_x )\delta (n_y )$ and infinite
media, the solution of this equation can be presented in the form:
\begin{eqnarray}
\hat {\rho }^{(0)}(\vec {k},z) = \hat {\rho }_{0} g(\vec {k},z),
\label{2.21}
\end{eqnarray}

\noindent where $\hat {\rho }_{0}$ is the spin part of the density
matrix [see  (\ref{1.5})]
\begin{eqnarray}\hat {\rho }_{0}= \frac{1}{3}I_{0}{\hat {\mbox I}} +
\frac{1}{2}\vec {P}_{0}\hat {\vec {S}} + \frac{1}{9}Q_{0ik}\hat
{Q}_{ik}, \label{2.22}
\end{eqnarray}

\noindent and the function
\begin{eqnarray} g(\vec {k},z)=\frac{1}{\pi \overline
{\,\theta ^2} z}\exp\left[{ - \frac{(\vec {n} - \vec {n}_0
)^2}{\overline {\,\theta ^2} z}}\right].\label{2.23}
\end{eqnarray}

\noindent The unit vector $\vec {n}=\vec{k}/k$ is counted off with
respect to  the direction of $\vec {n}_0 $ (further, we shall
align the $z$-axis with  vector $\vec {n}_0 )$.
 In the small-angle approximation,
$(\vec {n} - \vec {n}_0 )^2=\theta^{2}$, where $\theta$ is the
angle between vector $\vec{k}$ and the $z$-axis; $\overline
{\,\theta ^2}$ is the mean square scattering angle per unit path.
The mean square scattering angle ($\overline {\,\theta ^2} z$) at
depth $z$ will further be denoted by $\overline {\,\theta
^2_{z}}$, and its square root --- by ${\theta}_{z}$.

 We shall also
point out that for large values of $\theta $, the solution of
(\ref{2.16}) will generally be proportional to $1/\theta^{4} $
\cite{20, 21}. Thus, approximate equation (\ref{2.20}) reflects
the fact that the large--angle deflections are neglected in a
single scattering event. Indeed, for $\theta \gg \theta_{z} $, the
solution  (\ref{2.21}) decreases exponentially, while the
subsequent term that corresponds to the solution of the original
equation (\ref{2.16}) diminishes according to the power law
$\theta ^{ - 4}$. That is why the angular distribution
(\ref{2.23}) does not describe the characteristics of particles
scattered  at large angles, $\theta \gg \theta_{z}$ \cite{20, 21}.

 The expression for the density matrix in the zeroth
approximation (\ref{2.21}) enables one to obtain the correction
$\hat {\rho }^{(1)}(\vec {k},z)$ in the first--order perturbation
theory:
\begin{eqnarray}\fl
\hat {\rho }^{(1)}(\vec {k},z) = N\int_0^z {\int {G(\vec {k} -
{\vec {k}}'';z - z')} } \left( { - \sigma
_{\rm{NC}}^{\scriptscriptstyle 0} \hat {\rho }^{(0)}({\vec
{k}}'',z')  } \right.\nonumber
\\
-\frac{1}{2}\left( {\sigma _{\rm{NC}}^{\scriptscriptstyle\pm 1} -
\sigma _{\rm{NC}}^{\scriptscriptstyle0} }
\right)\left\{(\vec{S}\vec{n}'')^{2},\hat {\rho }^{(0)}({\vec
{k}}'',z') \right\}\nonumber
\\
+\left.{\frac{2\pi \rmi}{k}N \mathrm{Re} f_1 ^{\rm{nucl}}(0)
 \left[(\vec{S}\vec{n}'')^{2},\hat {\rho }^{(0)}({\vec {k}}'',z')
\right]} \right)\rmd\Omega _{\vec {k}''} \rmd z'\nonumber
\\
+ N\int_0^z {\int {G(\vec {k} - {\vec {k}}'';z - z' )} } \left(
{\hat {f}_{\rm{coul}} ({\vec {k}}'',{\vec {k}}')\hat {\rho
}^{(0)}({\vec {k}}',z' )\hat {f}^{\scriptscriptstyle +
}_{\rm{nucl,coul}} ({\vec {k}}',{\vec {k}}'') } \right. \nonumber
\\
+ \left. {\hat {f}_{\rm{nucl,coul}} ({\vec {k}}'',{\vec {k}}')\hat
{\rho }^{(0)}({\vec {k}}',z' )\hat {f}^{\scriptscriptstyle +
}_{\rm{coul}} ({\vec {k}}',{\vec {k}}'')} \right)\rmd\Omega _{\vec
{k}'} \rmd\Omega _{\vec {k}''} \rmd z'\nonumber
\\
+ N\int_0^z {\int {G(\vec {k} - {\vec {k}}'';z - z' )} }\nonumber
 \\
 \times\hat
{f}_{\rm{nucl,coul}} ({\vec {k}}'',{\vec {k}}')\hat {\rho
}^{(0)}({\vec {k}}',z' )\hat {f}^{\scriptscriptstyle +
}_{\rm{nucl,coul}} ({\vec {k}'},{\vec {k}''})\rmd\Omega _{\vec
{k}'} \rmd\Omega _{\vec {k}''} \rmd z' ,\label{2.24}
 \end{eqnarray}

\noindent where $G(\vec {k} - {\vec {k}}'';z - z' )$ is the Green
function of  (\ref{2.20}): $\displaystyle G(\vec {k} - {\vec
{k}}'';z - z' ) = \frac{1}{\pi \overline {\theta ^2} |z -
z'|}\exp\left[ - \frac{(\vec {n} - {\vec {n}}'')^2}{\overline
{\theta ^2} |z - z' |}\right].$ In the general case, the
amplitudes for Coulomb and nuclear scattering are given by formula
(\ref{1.6}).

 In the case of high energies, the characteristic angles for elastic
 scattering of deuterons by nuclei are $\theta_{\rm{n}}\sim1/kR_{\rm{d}}\ll
 1$ ($R_\rmd$ is the deuteron radius).
As a result, in analysing the polarization of high-energy deuteron
beams in transmitted geometry (\fref{f1}) using an axially
symmetrical detector, the amplitude of scattering due to nuclear
interaction, which appears in the expression for the density
matrix (\ref{2.24}), can be written in the form:
\begin{eqnarray}
\hat {f}_{\rm{nucl,coul}}(\vec {k},{\vec {k}}') = d(\theta ) + d_1
(\theta )(\vec {S}\vec {n}_0 )^2,\label{2.25}
\end{eqnarray}

\noindent where $\vec {n}_0$ is the unit vector directed along the
$z$-axis, which coincides with the direction of motion of the
initial beam; $d(\theta )$ and $d_1 (\theta )$ are the
spin--independent  and
 the spin--dependent parts of the nuclear
amplitude (which allows for the distortion caused to the waves
incident onto the nuclei by the Coulomb interaction),
respectively; $\theta$ is the angle between vectors $\vec {k}$ and
${\vec {k}}'$.

The spin structure of the Coulomb scattering amplitude $\hat
{f}_{\rm{coul}}$ also has the form  (\ref{2.25}). As stated above,
however, for this amplitude, the term
 proportional to the deuteron spin is small and will be dropped
 further.

 \subsubsection{Tensor Polarization of Deuterons Passing Through
Matter.}

By substituting the explicit form of the density matrix
(\ref{1.5}) and the scattering amplitude (\ref{2.25}) into the
solution (\ref{2.24}), one can obtain the characteristics of the
deuteron beam, which are of interest to us.

As a result, the number of particles that are registered by the
detector  with angular width $\Delta\Omega$ can be written in the
form:
\begin{eqnarray}\fl
{\cal I}(z,\vartheta_{\rm{det}}) = \left[1 - \exp\left( -
\frac{\vartheta _{\mathrm{det} }^2 }{\overline {\,\theta ^2_{z}}
}\right)\right]I_0 + \xi _1(z,\vartheta_{\rm{det}}) I_0 \nonumber
\\
+\left(\frac{}{}\xi _2 (z,\vartheta_{\rm{det}})+ \xi _3
(z,\vartheta_{\rm{det}})\right)\left[ {\frac{2}{3}I_0 +
\frac{1}{3}({\rm {\bf Q}}_0 \vec {n}_0 )\vec {n}_0 }
\right].\label{2.26}
 \end{eqnarray}

\noindent According to (\ref{2.26}), the intensity of the beam
that reaches
 the detector depends on its possible
initial tensor polarization ${\rm {\bf Q}}_0$.

The parameters $\xi _1 $, $\xi _2 $, and  $\xi _3$ in the
small-angle approximation are defined as follows:
\begin{eqnarray}
\xi _1 (z,\vartheta_{\rm{det}}) = - N\sigma
_{\rm{NC}}^{\scriptscriptstyle 0}\left[1 - \exp\left( -
\frac{\vartheta _{\rm{det} }^2 }{\overline {\theta
^2_{z}}}\right)\right]z\nonumber
\\
+ 2\pi N z\int\limits_0^\infty {P(\chi ;\vartheta _{\mathrm{\det}
} ,\overline {\,\theta ^2_{z}})\left( \frac{}{}{2 \mathrm{Re}
[a(\chi )d^ * (\chi )] + \vert d(\chi )\vert ^2} \right)} \chi
\rmd\chi,\nonumber
\\
 \xi _2 (z,\vartheta_{\rm{det}}) = - N(\sigma _{\rm{NC}}^{\scriptscriptstyle\pm 1} - \sigma _{\rm{NC}}^{\scriptscriptstyle 0} )\left[1 - \exp\left( -
\frac{\vartheta _{\mathrm{det} }^2 }{\overline {\theta
^2_{z}}}\right)\right]z\nonumber
\\
+ 4\pi Nz\int\limits_0^\infty {P(\chi ;\vartheta _{\rm\det }
,\overline {\,\theta ^2_{z}})\left(\frac{}{}\mathrm{Re}[a(\chi
)d_1^ * (\chi )]\right.}+
 \left.{\mathrm{Re}[d(\chi
)d_1^
* (\chi )]}\frac{}{} \right)
\chi \rmd\chi,\nonumber
\\
\xi _3 (z,\vartheta_{\rm{det}}) = 2\pi Nz\int\limits_0^\infty
{P(\chi ;\vartheta _{\rm\det } ,\overline {\,\theta ^2_{z}})\vert
d_1 (\chi )\vert ^2} \chi \rmd\chi,\label{2.27}
 \end{eqnarray}

\noindent where the function $P(\chi ;\vartheta _{\rm\det }
,\overline {\,\theta ^2_{z}})$ is defined as
\begin{eqnarray}
P(\chi ;\vartheta _{\mathrm{det} } ,\overline {\,\theta ^2_{z}}) =
\frac{2}{\overline {\,\theta ^2_{z}}}\int\limits_0^{\vartheta
_{\mathrm{det} } } {\exp\left( - \frac{\theta ^2 + \chi
^2}{\overline {\,\theta ^2_{z}}}\right)\rm{I}_0 \left(
{\frac{2\theta \chi }{\overline {\,\theta ^2_{z}}}} \right)\theta
\rmd\theta },\label{2.28}
\end{eqnarray}
where $\rm{I}_0 (y)$ is the modified Bessel function of the zeroth
order.

Let us discuss now what polarization this beam has. Write the
expressions for the number of particles that are registered by the
detector and have the initial spin projections $M=0$ and $M=\pm
1$, respectively:
\begin{eqnarray}
 {\cal I}^{\scriptscriptstyle(0)}(z,\vartheta_{\rm{det}})
 &=&
  \left[1 - \exp\left( - \frac{\vartheta _{\rm{det} }^2 }{\overline {\,\theta ^2_{z}}
}\right)\right]I_0^{\scriptscriptstyle(0)} + \xi _1
I_0^{\scriptscriptstyle(0)},\nonumber
\\
 {\cal I}^{\scriptscriptstyle(\pm1)}(z,\vartheta_{\rm{det}})
 &=&
  \left[1 - \exp\left( - \frac{\vartheta _{\mathrm{det} }^2 }{\overline {\,\theta ^2_{z}}
}\right)\right]I_0^{\scriptscriptstyle(\pm1)} + (\xi _1+\xi _2 +
\xi _3 ) I_0^{\scriptscriptstyle(\pm1)}\label{2.29}
 \end{eqnarray}

For the sake of simplicity, let us consider the case when an
unpolarized beam is incident onto the target. Using the definition
(\ref{2.3}), find tensor polarization of the beam registered by
the detector:
\begin{eqnarray}
 p_{zz}\simeq\frac{2}{3\left[1 - \exp\left( - \frac{\vartheta
_{\mathrm{det} }^2 }{\overline {\,\theta ^2_{z}}
}\right)\right]}(\xi _2 + \xi _3).\label{2.30}
\end{eqnarray}

Here we have retained the terms proportional to the first power of
the small quantities $\xi _2$ and $\xi _3$.

Using the explicit form of the parameters $\xi _2$, $\xi _3$ in
(\ref{2.27}), the expression for tensor polarization of the beam
can be written in the form:
\begin{eqnarray}\fl
p_{zz}(z, \vartheta_{\rm{det}})=-\frac{2}{3}N(\sigma
_{\rm{r}}^{\scriptscriptstyle\pm 1} - \sigma
_{\rm{r}}^{\scriptscriptstyle 0} )z -\frac{2}{3}N\Bigg[(\sigma
_{\rm{el}}^{\scriptscriptstyle\pm 1} - \sigma
_{\rm{el}}^{\scriptscriptstyle 0} )\nonumber
\\
-\frac{4\pi}{\left[1 - \exp\left( - \frac{\vartheta _{\rm{det} }^2
}{\overline {\,\theta ^2_{z}} }\right)\right]}
\int\limits_0^\infty P(\chi ,\vartheta _{\det }, \overline
{\,\theta ^2_{z}})\frac{}{}\mathrm{Re}[a(\chi )d_1^ * (\chi )]\chi
\rmd\chi\nonumber
\\
 -\frac{2\pi}{\left[1 - \exp\left( - \frac{\vartheta _{\rm{det} }^2
}{\overline {\,\theta ^2_{z}} }\right)\right]}
 \int\limits_0^\infty P(\chi ,\vartheta _{\det }, \overline
{\,\theta ^2_{z}})\nonumber
\\
 \times\left.\left(2\mathrm{Re}[d(\chi )d_1^ * (\chi
)]+|d_{1}(\chi)|^{2}\right)\chi
d\chi\frac{}{}\right]z.\label{2.31}
\end{eqnarray}

Equation (\ref{2.31}) simplifies appreciably  for the two limiting
values of the detector angle: $\vartheta_{\rm{det}}\ll \theta_{z}$
and $\vartheta_{\rm{det}}\gg \theta_z$. As has been stated above,
the solution (\ref{2.21}) does not describe the properties of the
beam that is scattered at  angles $\theta
>\theta_{z}$. But the integral characteristics of deuterons that are obtained
from this solution can be used for
$\vartheta_{\mathrm{det}}>\theta_{z}$. The influence of particles
single--scattered due to the Coulomb interaction can be neglected
because the major contribution to (\ref{2.28}) comes from the
small values of $\theta$.

When $\vartheta_{\rm{det}}$ is much less than $\theta_z$, the
expression for $p_{zz}$ can be written in the form:
\begin{eqnarray}\fl
p_{zz}(z, \vartheta_{\mathrm{det}}\ll \theta_{z})=
-\frac{2}{3}N(\sigma _{\rm{tot}}^{\scriptscriptstyle\pm 1} -
\sigma _{\rm{tot}}^{\scriptscriptstyle 0} )z\nonumber
\\
+\frac{8}{3}\pi N z \int\limits_0^\infty
\exp\left(-\frac{\chi^{2}}{\overline {\,\theta
^2_{z}}}\right)\frac{}{}\mathrm{Re}[a(\chi )d_1^ * (\chi )]\chi
\rmd\chi\nonumber
\\
 +\frac{4}{3}
\pi N z
 \int\limits_0^\infty \exp\left({-\frac{\chi^{2}}{\overline {\,\theta ^2_{z}}}}\right)\left(2\mathrm{Re}[d(\chi )d_1^ * (\chi
)]+|d_{1}(\chi)|^{2}\right)\chi \rmd\chi.\label{2.32}
\end{eqnarray}

According to (\ref{2.32}), as a result of multiple scattering,
tensor polarization  becomes independent of
$\vartheta_{\mathrm{det}}$ for small angles of the detector.
 Equation (\ref{2.32}) contains a scattering-related contribution
(the second and third terms), which appears in (\ref{2.32}) as a
result of deuteron rescattering from the direction of
$\vec{k}\neq\vec{k}_{0}$ into the direction of vector
$\vec{k}_{0}$.

In the other limiting case, when $\vartheta_{\rm{det}}$ is much
greater than $\theta_z$, the terms related to scattering in
(\ref{2.31}) are equal to the scattering cross sections, and so
the expression between the square brackets vanishes. Tensor
polarization is determined by the difference between the cross
sections $\sigma_{\rm{r}}^{\pm1}$ and $\sigma^0_{\rm{r}}$ alone,
as it occurs in the case of a thin target.

Let us represent the expression (\ref{2.31}) for tensor
polarization in the form:
\begin{eqnarray}
p_{zz}(z, \vartheta_{\rm{det}})=-\frac{2}{3}N(\sigma
_{\rm{r}}^{\scriptscriptstyle\pm 1} - \sigma
_{\rm{r}}^{\scriptscriptstyle 0} )z-\frac{2}{3}N(\sigma
_{\rm{el}}^{\scriptscriptstyle\pm 1} - \sigma
_{\rm{el}}^{\scriptscriptstyle 0} )z\,\mathrm{H}(z,
\vartheta_{\mathrm{det}}),\label{2.33}
\end{eqnarray}

\noindent where the function $\mathrm{H}$ is defined as
\begin{eqnarray}\fl
\mathrm{H}(z, \vartheta_{\mathrm{det}})=1-\frac{1}{(\sigma
_{\rm{el}}^{\scriptscriptstyle\pm 1} - \sigma
_{\rm{el}}^{\scriptscriptstyle 0} )}\frac{4\pi}{\left[1 -
\exp\left({ - \frac{\vartheta _{\rm{det} }^2 }{\overline {\,\theta
^2_{z}} }}\right)\right]} \left(\int\limits_0^\infty P(\chi
,\vartheta _{\det }, \overline {\,\theta
^2_{z}})\frac{}{}\mathrm{Re}[a(\chi )d_1^ * (\chi )]\chi
\rmd\chi\right.\nonumber
\\
 +\left.
 \int\limits_0^\infty P(\chi ,\vartheta _{\det }, \overline
{\,\theta ^2_{z}})\left(\mathrm{Re}[d(\chi )d_1^ * (\chi
)]+|d_{1}(\chi)|^{2}/2\right)\chi \rmd\chi\right).\label{2.34}
\end{eqnarray}

Note that when passing to the limit of a thin target, i.e.,
$z\rightarrow0$, one should take into account that in this case,
the function $P(\chi ;\vartheta _{\det } ,\overline {\,\theta
^2_{z}})$ turns to $\displaystyle P(\chi ;\vartheta _{\rm{det} }
,0) = \int\limits_0^{\vartheta _{\mathrm{det} } } {\delta
(\vartheta - \chi )\rmd\vartheta } $, where $\delta$ is the delta
function.

According to (\ref{2.33}), the dependence of tensor polarization
of a beam on the target thickness and the angle $\vartheta
_{\mathrm{det} }$ is determined by the function
$\rm{H}(z,\vartheta _{\mathrm{det} })$.

Let us recast the equality for the mean square angle of deuteron
scattering at depth $z$ in the form \cite{19}:
\begin{eqnarray}
\displaystyle \displaystyle\overline {\theta ^2}z = 16\pi
Z^2\left( {\frac{e^2}{pv}} \right)^2\mbox{ln}(137\, Z
^{-1/3})\frac{N_{A}}{A}\Delta C,\label{2.35}
\end{eqnarray}

\noindent where  $\Delta C$ is the target thickness in g/cm$^2$,
$N_{A}$ is the Avogadro's number.

The explicit form of  the $d$ dependence on $\theta $  in the
range of small scattering angles for the case of structural
particles was derived, for example, in \cite{22}. To estimate the
main parameters, let us consider the simplest form of the $d$ and
$d_1 $ dependence on $\theta $, namely,
\begin{eqnarray}
d(\theta ) = d(0)\exp\left( - k^2R_{\rmd}^2 \theta ^2 /
4\right),\qquad d_1 (\theta ) = d_1 (0)\exp\left( - k^2R_{\rmd}^2
\theta ^2 / 4\right),\label{2.36}
\end{eqnarray}

\noindent where $R_{\rmd} $ is the deuteron radius.

We also used the fact that according to the eikonal-approximation
calculations \cite{23}, in the range of energies of incident
deuterons from 400 to 800 MeV, for the amplitude $d$ we have
$\mbox{Im}d = 0.75 \cdot 10^{ - 11}$cm, $\mbox{Re}d = - 0.6 \cdot
10^{ - 12}$cm, and the  imaginary part of the amplitude $d_1$
relates to its real part as $\mathrm{Im}d_{1}/\mathrm{Re}d_{1}\sim
-10$,  with the spinless part of the nuclear amplitude
$f_{\rm{nucl}}$ being much larger than its spin part.

For the considered  energies of  deuterons scattered by a screened
Coulomb potential,  the first Born approximation to the amplitude
is satisfactory to use as a real part of the Coulomb amplitude
$a(\theta)$, and the second term of the Born series can be used as
its imaginary part accordingly.

\begin{figure}[!h]
\centering
\includegraphics[scale=0.9]{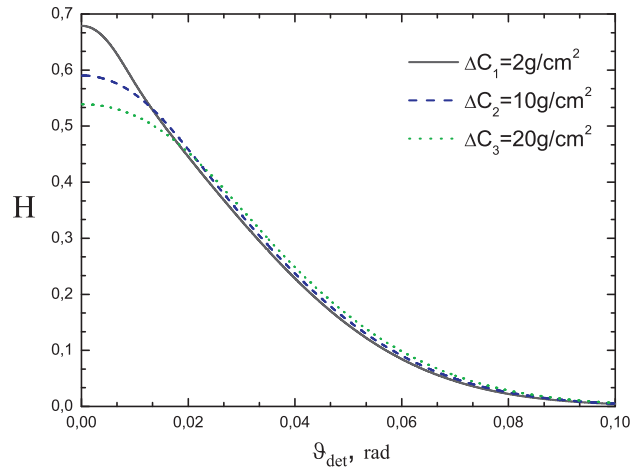}
\caption{Function $\mathrm{H}$ versus the detector angle for
deuteron energy of 500 MeV. Solid curve corresponds to the carbon
target thickness $\Delta C_{1}=2$ g/cm$^{2}$, dashed curve - to
$\Delta C_{2}=10$ g/cm$^{2}$, dotted curve --- to $\Delta
C_{3}=20$ g/cm$^{2}$.}\label{f2}
\end{figure}

\begin{figure}[!h]
\centering
\includegraphics[scale=0.9]{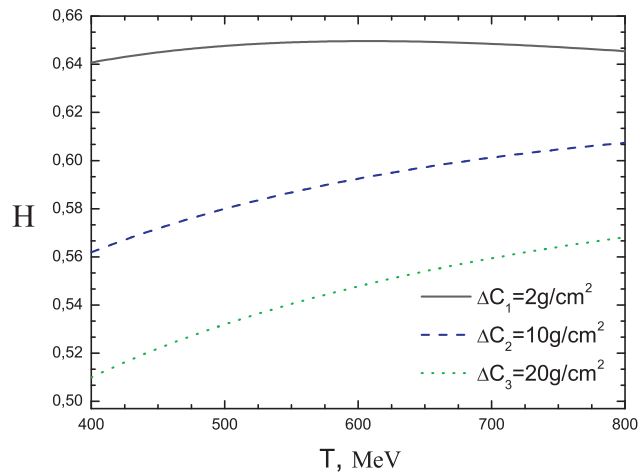}
\caption{Function $\mathrm{H}$ versus the deuteron energy  for the
detector angle $\vartheta_{\mathrm{det}}=5\cdot10^{-3}$ rad. Solid
curve corresponds to the carbon target thickness  $\Delta C_{1}=2$
g/cm$^{2}$, dashed curve --- to $\Delta C_{2}=10$ g/cm$^{2}$,
dotted curve --- to $\Delta C_{3}=20$ g/cm$^{2}$.}\label{f3}
\end{figure}

\Fref{f2} and \fref{f3} present the results of calculations of the
function $\rm{H}$ for different values of $\vartheta
_{\mathrm{det} }$ and different  deuteron energies {when a beam is
scattered by a carbon filter.}

The results of calculations given in \fref{f2} demonstrate that
owing to multiple Coulomb scattering, the   value of $\mathrm{H}$
is less than unity, $\mathrm{H}<1$, even in a narrow--angle
geometry ($\vartheta_{\rm{det}}\ll \theta_{z}$). Let us recall
that this occurs because of the  rescattering in the direction of
vector $\vec{k}_{0}$ of  particles that have been scattered in
different directions. For $\Delta C=2,\, 10$, and $20$ g/cm$^{2}$,
the values of $\theta_{z}=\sqrt{\overline{{\theta^{2}}}z}$ are
$\theta_{z}(\Delta C_{1})=4.5$~mrad, $\theta_{z}(\Delta
C_{2})=10$~mrad, and $\theta_{z}(\Delta C_{3})=14$~mrad,
respectively. As a result, in this limiting case,  the values of
$\mathrm{H}$ for deuterons registered by the detector with angular
width $\vartheta_{\rm{det}}\ll \theta_{z}$ for target thicknesses
$\Delta C=2,\, 10$, and  $20$ g/cm$^{2}$ are $\mathrm{H}(\Delta
C_{1})=0.68$, $\mathrm{H}(\Delta C_{2})=0.59$, and
$\mathrm{H}(\Delta C_{3})=0.54$, respectively. As is seen, these
values differ from unity by almost a factor of two, and this
difference is the largest for the thickest target.

With growing angle $\vartheta_{\rm{det}}$, the particles scattered
at the angles $\theta\leq \vartheta_{\rm{det}}$ gain in
significance, which leads to a decrease of $\mathrm{H}$. In a
wide--angle geometry, which for the fast--particle scattering is
defined from the condition
$\vartheta_{\rm{det}}\gg\theta_{\rm{eff}}$
($\theta_{\rm{eff}}=\theta_{z}$ when $\theta_{z}>\theta_{\rm{n}}$,
and $\theta_{\rm{eff}}=\theta_{\rm{n}}$  when
$\theta_{z}<\theta_{\rm{n}}$), we have
 $\mathrm{H}\rightarrow 0$.  As a result, tensor polarization in this limiting case is
 determined by the difference between the inelastic parts of the
 total scattering cross sections  for the   states with
$M=\pm1$ and $M=0$.

Now, let us consider the behaviour of the function $\rm{H}$ for
different values of energy at a fixed value of
$\vartheta_{\mathrm{det}}$ (choose $\vartheta_{\rm{det}}$ to be
$\vartheta_{\rm{det}}=5\cdot10^{-3}$~rad). According to \fref{f3},
when the deuteron energy grows,  the value of $\mathrm{H}$  first
increases and then starts  to diminish.

It should be emphasized that with due account of multiple Coulomb
scattering, tensor polarization is a nonlinear function of the
target thickness $z$ (\ref{2.31}). For large values of
$\vartheta_{\rm{det}}$, however,  $p_{zz}$ becomes a linear
function of $z$ (\ref{2.10}). With increasing $\overline {\,\theta
^2_{z}}$, expression  (\ref{2.32}) approaches  the expression for
tensor polarization (\ref{2.10}). This means that for large values
of $z$ or for low energies of the incident deuterons, the
dependence of the tensor polarization on the detector's angle
becomes insignificant  (it is necessary that the parameter
$\theta_{z}$ should be much greater than the characteristic angle
$\theta_{\rm{n}}$ of nuclear scattering). In this case, tensor
polarization can be considered linear in $z$ for any values of
$\vartheta_{\rm{det}}$.

\section{Conclusion}

It has been shown that the  magnitude of tensor polarization of
the deuteron beam, which arises from the spin dichroism effect,
depends appreciably on the angular width $\vartheta_{\rm{det}}$ of
the detector that registers the deuterons transmitted through the
target. Even when the angular width of the detector is much less
than the mean square angle of multiple Coulomb scattering, the
beam's tensor polarization depends noticeably on rescattering. In
the case when the angle $\vartheta_{\rm{det}}$ is much larger than
the mean square angle of multiple Coulomb scattering (as well as
than the characteristic angle of elastic nuclear scattering),
tensor polarization is determined by the difference between the
total
 reaction cross sections for deuteron--nucleus interaction,
$\sigma_{\rm{r}}^{\pm 1}-\sigma^0_{\rm{r}}$. Elastic scattering
processes here make no contribution to tensor polarization.

\section*{Acknowledgements}

The authors are very grateful to A. Rouba for calculating the
forward nuclear amplitudes.


\section*{References}
\begin{thebibliography}{22}

\bibitem{1} Baryshevsky V G 1992 {\it Phys. Lett.} A {\bf 171} 431--34

\bibitem{2} Baryshevsky V G 1993 {\it J. Phys.} G {\bf 19} 273--82

\bibitem{3} Baryshevsky  V G  2012 {\it High-Energy Nuclear Optics of Polarized
Particles}  (Singapore: World Scientific)

\bibitem{4} Baryshevsky V  \etal 2005 First observation of spin dichroism with deuterons up
to 20 MeV in a carbon target {\it Preprint} hep-ex/0501045

\bibitem{5} Baryshevsky V \etal 2007
{\it Proc. of the 17th int. spin physics symp. (Kyoto)}  (New
York: Melville) p 777

\bibitem{6} Seyfarth H  \etal 2011 {\it Phys.
Rev. Lett.} {\bf 104}  222501

\bibitem{7} Azhgirey L S  \etal 2008 {\it Proc. of the XIIth  Work. on high energy spin
physics (Dubna)} (Dubna: JINR) p 205


\bibitem{8} Azhgirey L S \etal 2010
{\it Phys. Part. Nuclei Lett.} {\bf 7}  27--32

\bibitem{9} Rathmann F and  Nikolaev N 2011 {\it Proc. of the 8th
Int. Conf. on Nuclear Physics at Storage Rings (Frascati)} (Italy
PoS(STORI11)) p 029

\bibitem{10} Baryshevsky V G 2008  {\it J. Phys.} G {\bf 35}   035102


\bibitem{11} Baryshevsky V G and  Rouba A 2010 {\it Phys. Lett.} B
{\bf 683}  229--34

\bibitem{12} Baryshevsky V G and  Shyrvel A R {\it Preprint} hep-ph/1101.2408v1

\bibitem{13} Baryshevsky V G and  Shekhtman A G 1996 {\it Phys.
Rev.} C {\bf 53}  267--76.

\bibitem{14} Goldberger  M L and  Watson K M 1964 {\it Collision Theory} (New York: John
Wiley and Sons)

\bibitem{15}  Ohlsen G G 1972 {\it Rep. Prog. Phys.} {\bf 35}
717--801.


\bibitem{16} Darden S E 1967 {\it Am. J. Phys.} {\bf 35}   727--38.

\bibitem{17}  Davydov A S 1965 {\it Quantum Mechanics} (Oxford: Pergamon
Press)

\bibitem{18} Lyuboshitz V L 1980  {\it Sov. J. Nucl. Phys.} {\bf 32}
 362--65.

\bibitem{19}  Ter-Mikaelian M  L 1972 {\it High Energy Electromagnetic Processes in
Condensed Media} (New York: Wiley Inter-science)

\bibitem{20} Bethe H A  1953 {\it Phys. Rev.}   {\bf 89}  1256--70

\bibitem{21} Moliere G 1948 {\it Z. Naturforsch} A {\bf 3}  78--96
\bibitem{22} Czyz W and  Maximon L C 1969 {\it Ann. Phys.} (N.Y.) {\bf 52}  59--121.
\bibitem{23} Rouba A Private Communication
\end{thebibliography}
\end{document}